# Raman-free, noble-gas-filled PCF source for ultrafast, very bright twin-beam squeezed vacuum

**Martin A. Finger[1*], Timur Sh. Iskhakov[1†], Nicolas Y. Joly[2,1], Maria V. Chekhova[1,2,3] and Philip St.J. Russell[1,2]**

[1]*Max Planck Institute for the Science of Light and* [2]*Department of Physics, University of Erlangen-Nuremberg, Guenther-Scharowsky Strasse 1, Bau 24, 91058 Erlangen, Germany*

[3]*Physics Department, Lomonosov Moscow State University, Moscow 119991, Russia*

*e-mail: martin.finger@mpl.mpg.de*

We report a novel source of twin beams based on modulational instability in high-pressure argon-filled hollow-core kagomé-style photonic-crystal fibre. The source is Raman-free and manifests strong photon-number correlations for femtosecond pulses of squeezed vacuum with a record brightness of ~2500 photons per mode. The ultra-broadband (~50 THz) twin beams are frequency tunable and contain one spatial and less than 5 frequency modes. The presented source outperforms all previously reported squeezed-vacuum twin-beam sources in terms of brightness and low mode content.

---

[†] Now working at Technical University of Denmark (DTU)

Correlated photons and twin beams, among very few accessible nonclassical states of light, are at the focus of modern quantum optics. Their applications in quantum metrology [1-4], imaging [5-7], key distribution [8] and other fields make them the basic resource in photonic quantum technologies. Entangled photon pairs can be generated by parametric down-conversion (PDC) or four-wave mixing (FWM) sources. Depending on the pumping strength, the state emerging at the output of the nonlinear material is called squeezed vacuum (SV) at low photon flux [9], or bright squeezed vacuum (BSV) at high photon flux [10]. The mean field of such states is zero, while the mean energy can approach high values.

In twin-beam SV, there is strong correlation in the photon numbers emitted into the two conjugated beams (called signal and idler). At low photon flux, this simply means that the emitted photons come rarely but always in pairs. For BSV, the numbers of photons emitted into signal and idler beams are very uncertain but always exactly the same. The standard technique for detecting photon-number correlations in this case is to measure the noise reduction factor (*NRF*), which is the variance of the photon-number difference between the signal and idler channels, normalized to the shot-noise level, i.e., the mean value of the total photon number [11]:

$$NRF = \frac{Var(N_s - N_i)}{\langle N_s + N_i \rangle}, \quad (1)$$

where $N_s$, $N_i$ are the photon numbers in the signal and idler modes. *NRF* is a measure of the reduction of quantum noise below the shot-noise level, which corresponds to *NRF* = 1.

Fibre-based SV sources are complementary to crystal-based sources and especially interesting because they allow one to engineer temporal correlations while eliminating spatial correlations [12]. The wide flexibility in designing the time-frequency mode structure allows high dimensional temporal Hilbert spaces to be exploited, making such systems promising candidates for more secure quantum cryptography [13] or for high-information-capacity quantum communications, by analogy with multimode states in space [14]. Temporally few-mode-systems are a particularly interesting resource in quantum optics, even though addressing and selecting individual frequency modes still presents a challenge [15, 16]. Apart from this, fibre-based sources are attractive because of easy manufacturing, high conversion efficiencies due to long optical path-lengths, and integrability into optical networks. A major design concern for fibred SV sources is, however, the unavoidable Raman noise [17, 18, 19, 20]. There have been many attempts to reduce the deleterious effects of spontaneous Raman scattering (SpRS) on photon correlations, examples being cryogenic cooling of the fibre to 4 K [21], the use of cross-polarized phase-matching in birefringent fibres [19] or employing crystalline materials [22]. These techniques suffer, however, from technological difficulties, incomplete suppression of SpRS or large coupling losses. Another way to reduce SpRS is to generate twin beams spectrally well separated from the pump wavelength. This can be achieved by pumping the fibre in the normal dispersion region, where the generating nonlinear process is FWM [23]. In this regime, the deleterious effects of SpRS are strongly reduced because the sideband spacing is much greater than the Raman shift. Nevertheless, higher-order Raman scattering still corrupts photon-number correlations [18]. Additionally, the wide sideband spacing prevents femtosecond-pulse pumping because signal, idler, and pump photons suffer strong group velocity walk-off [18]. In contrast, pumping the fibre in the anomalous dispersion regime leads to modulational instability (MI), which in the presence of the optical Kerr effect can be phase-matched. Under these conditions the signal and idler bands lie very close to the pump frequency, at the same time being broad (Fig. 1b) and power-dependent (see Supplementary Section 2). Their proximity to the pump makes them normally difficult to use due to degradation of the lower-frequency sideband by photons Raman-scattered from the pump [18, 19, 20].

The generation of SV by FWM or MI is described by a Hamiltonian similar to the one for PDC (see Supplementary Section 1). In such a system the mean number of photons per mode is given by $\langle N \rangle = \sinh^2(G)$, assuming the pump is undepleted [24]. The parametric gain $G$ is proportional to the pump power, the optical path-length in the nonlinear material and the $\chi^{(3)}$ nonlinearity [25]. Despite originating from the same fundamental mechanism, the properties of FWM and MI are substantially different, justifying distinguishing the two processes [18].

In this letter, we present the first application of gas-filled hollow-core photonic-crystal fibre (PCF) for generating nonclassical states of light. Nonlinearity is provided by argon, which intrinsically avoids SpRS due to its monatomic structure. The fibre is a kagomé-lattice hollow-core, which has all the essential ingredients for generating ultrafast twin beams. It offers transmission windows several hundred nm wide at moderate losses (typically a few dB/m or less), combined with small values of anomalous group velocity dispersion (GVD) that are only weakly wavelength-dependent [26] and can be balanced against the normal dispersion of the filling gas, giving rise to a pressure-dependent zero-dispersion wavelength (ZDW) [27]. This results in turn in large tunability of the signal and idler wavelengths, something that is impossible in solid-core fibre systems even with the help of power-dependent phase-matching in the MI regime (see Supplementary Section 2) [18]. Furthermore, the weak wavelength-dependence of the dispersion allows generation of ultra-broadband MI sidebands with widths greater than 50 THz (Fig. 1b).

In the setup (Fig. 1a), an argon-filled 30-cm-long kagomé-PCF with a core diameter of 18.5 µm (flat-to-flat) and a core wall thickness of 240 nm was pumped by an amplified Ti:Sa laser at 800 nm with pulse rate 250 kHz and duration ~300 fs (after a 3.1 nm bandpass filter). Based on an empirical model for the dispersion of kagomé-PCF (*s*-parameter = 0.03 [28]), these parameters yield a ZDW at 770 nm for a pressure of 75 bar. The pulse energy in the fibre core varied between 140 and 350 nJ, and polarization optics in front of the PCF allowed linear polarized light to be aligned along one eigen-axis of the weakly birefringent fibre. After exiting the PCF the light was collimated with an achromatic lens and the sidebands separated from the pump beam by spectral and spatial filtering. This yielded an overall pump suppression of at least 95 dB. The two sidebands were finally detected using standard silicon PIN photodiodes from Hamamatsu (S3759 and S3399) with ~95% quantum efficiency (protection windows removed), followed by charge sensitive amplifiers, which generate voltage pulses whose area scales with the number of photons per optical pulse. The amplification factors of the detectors were measured to be (2.096±0.002) pV.s/photon for the detector in the high frequency sideband and (2.178±0.004) pV.s/photon for the detector in the low frequency sideband. The noise of the photodetectors was determined from the standard deviation of the signals without light. The detector noise amounted to ~600 photons/pulse for the short wavelength sideband and ~650 photons/pulse for the long wavelength sideband. In order to account for the detector response we calibrated the shot-noise level of the system to the coherent state of the laser (see Supplementary Section 3) [29].

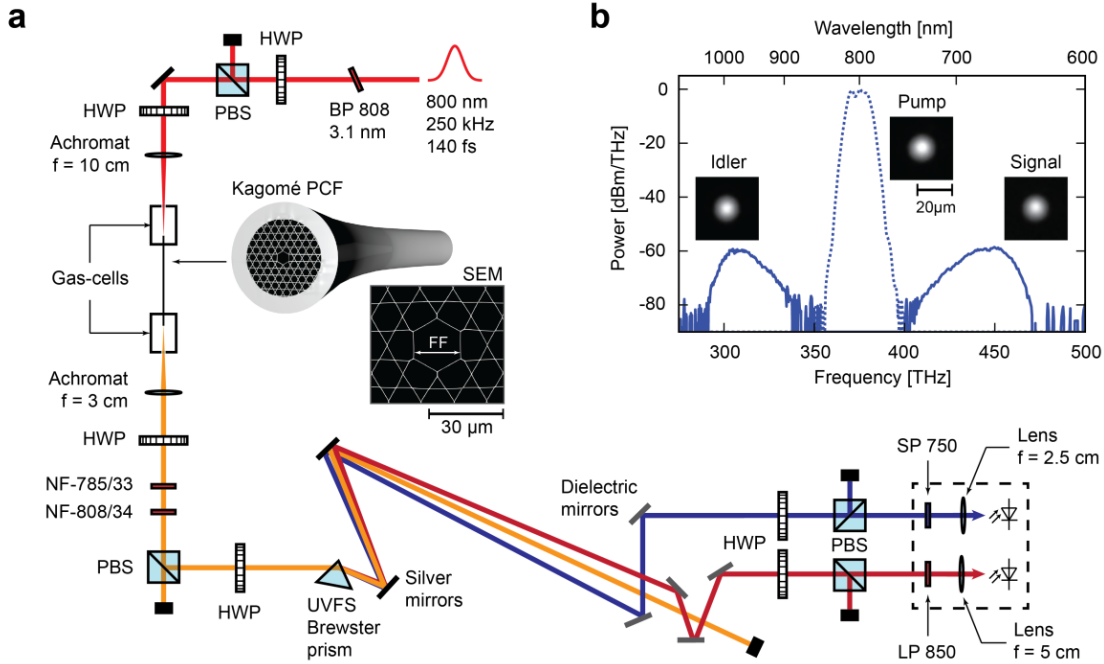

FIG. 1 (color). (a) Setup for twin beam generation and analysis. BP = bandpass filter, HWP = half-wave plate, PBS = polarizing beam splitter, NF = notch filter, LP = long-pass filter, SP = short-pass filter, SEM = scanning electron micrograph, FF = flat-to-flat diameter. (b) Modulation instability spectrum at 75 bar and 250 nJ pulse energy measured with an optical spectrum analyser (OSA / ANDO AQ 6315-E). The dotted curve shows the unfiltered spectrum, while the solid curve shows the spectrum after blocking the pump light with notch filters. The insets show near-field mode profiles, confirming that the light is in the fundamental mode.

The spectral location of the sidebands was widely tunable by changing the pressure and therefore shifting the ZDW. The overall accessible spectral range (ASR) of the MI sidebands depended on pump power, dispersion (i.e. pressure) and pulse duration. Adjusting the filling pressure from 50 to 96 bar, we were able to tune each sideband by ~80 THz at a fixed pulse energy of 320 nJ, corresponding to a wavelength shift of ~200 nm on the infrared side (see Fig. 2a). Since the pump powers at which we observed nonclassical noise reduction are too small to measure the corresponding spectra directly, we performed nonlinear pulse propagation simulations based on the generalized nonlinear Schrödinger equation. Figure 2b shows the pressure dependence of the spectrum at a parametric gain of 4.2, which is in the middle of the parametric gain interval used to observe noise reduction in the system. Note that the ASR limits of the NRF measurement are marked in the figure. Ultimately, the spectral range is limited by the fading nonlinearity at low pressures and by a strongly reduced gain at high pressures due to a larger group-velocity mismatch between the sidebands and the pump. In our case, any spectral cut-off due to group-velocity walk-off is masked by a loss peak in fibre, caused by an anti-crossing between the core mode and a cladding resonance (Fig. 2b).

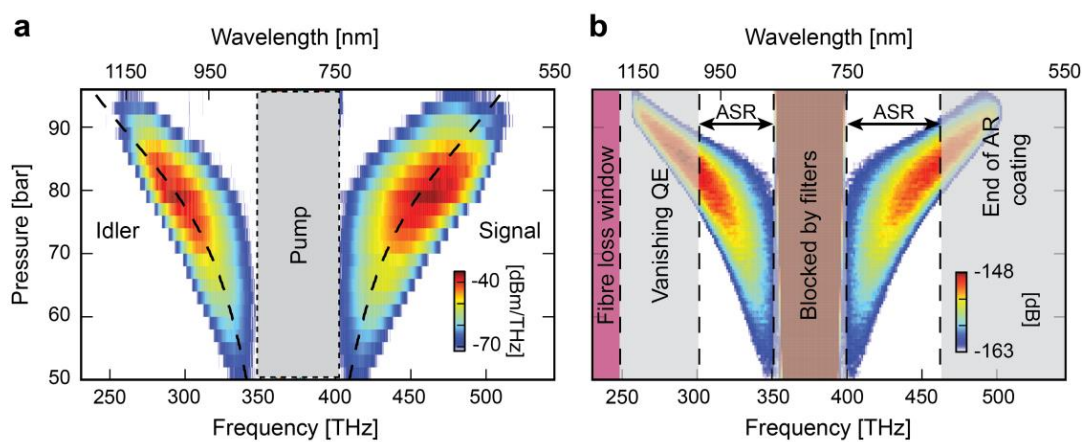

FIG. 2 (color). (a) Measured pressure dependence of the MI sidebands for a constant pump pulse energy of 320 nJ. The spectra were measured with the pump blocked by notch filters. The dashed line marks the locus predicted for perfect phase-matching. (b) Simulation of the pressure dependence of the nonclassical light spectra at a parametric gain of 4.2. The accessible spectral range (ASR) for measuring twin-beam squeezing is limited by the quantum efficiency (QE) of the silicon photodiode and by the width of the antireflection (AR) coating on the optical components.

Figure 3a shows the dependence of *NRF* on the sum of the mean numbers of detected signal and idler photons at three different pressures. It can be seen that the measured noise in the photon-number difference is ~35% below shot noise, indicating nonclassical correlation of photon numbers between signal and idler beams. The twin-beam squeezing is observed up to a record value (for squeezed vacuum) of ~2500 photons per mode – almost three times brighter than crystal-based sources, which have demonstrated squeezing up to ~900 photons per mode [11]. The measured *NRF* is in good agreement with loss estimates in the optical channel (see Supplementary Section 4). The parametric gain was varied within the range 3.9 to 4.6 by increasing the pump power. When the pressure was changed from 70 bar to 76 bar and the pump pulse energy was kept constant (180 nJ), the signal and idler bands shifted away by ~22 THz. The change in slope for different pressures can be explained by contributions from distinct unmatched modes [11], caused by the unequal frequency dependence of the detection channels. When the number of the unmatched modes increases, the slope becomes steeper due to more uncompensated intensity fluctuations. The accessible pressure range for observation of twin-beam squeezing is limited by a sharp drop-off in the quantum efficiency of the idler silicon photodetector above 1000 nm (see Fig. 2b). In the current configuration, we found that the best working pressure to be ~76 bar.

Next we measured the dependence of *NRF* on loss. In contrast to the normalized Glauber correlation functions, *NRF* is highly sensitive to losses in the optical channels, the best possible value being $1 - \eta$, where $\eta$ is the quantum efficiency of the optical channel after the PCF [30]. This was tested by monitoring *NRF* while increasing the loss symmetrically in both channels by adjusting the half-wave plate after the fibre. A linear increase in *NRF* up to the SNL is observed (Fig. 3b), in good agreement with theory. The measured *NRF* can be further improved by reducing the loss in the system, e.g., the loss of the kagomé-PCF (lowest loss reported at 800 nm is 70 dB/km [31]). With such state-of-the-art PCF and antireflection-coated gas-cell windows, the source itself would introduce a total loss of ~1.5%. This would correspond to an optimal *NRF* of −18 dB; all other losses relate to the detectors and are the same for any source.

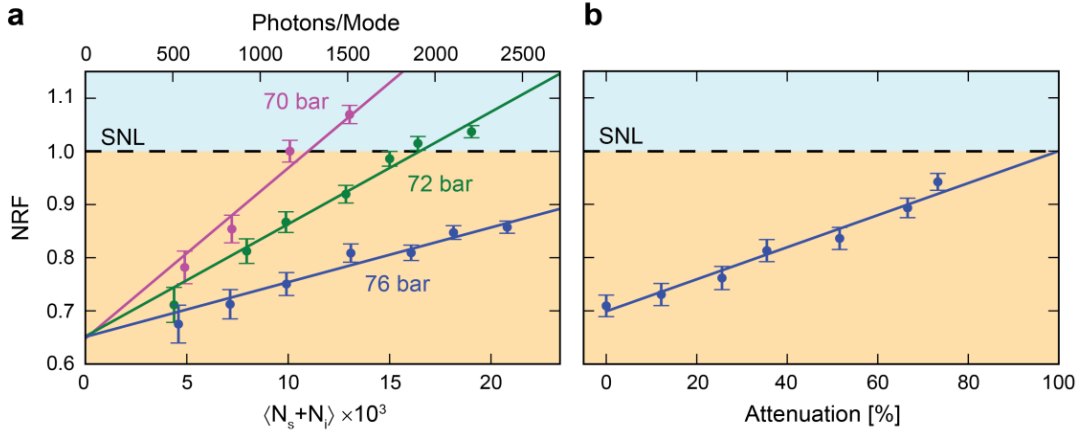

FIG 3. (color). (a) *NRF* plotted against the average total number of twin-beam photons, changed by varying the pump power, at 70, 72 and 76 bar. The difference in slope is caused by unequal frequency dependence of the detection efficiency in the two optical channels (the sideband frequencies shift when the pressure is changed). (b) Measured values of *NRF* plotted against attenuation in the optical channel performed with the HWP and the PBS. Each data point in the diagram corresponds to one million recorded pulses. Error bars represent the standard error of the measurement.

We determined the effective number of spatiotemporal modes K in the system by measuring the second-order intensity correlation function $g^{(2)}$ (see Supplementary Section 5) [32], corresponding to the product of effective spatial and temporal mode numbers. Here, we measured $g^{(2)} = 1.232\pm0.014$, which leads to $K = 4.31\pm0.26$. By inserting a 10 nm bandpass filter centred at 960 nm into the long-wavelength channel, the correlation time of the intensity fluctuations was increased, allowing us to measure only the number of spatial modes, which turned out to be 1. This means that the measured number of modes is equal to the effective number of temporal modes. The spatially single-mode behaviour, confirmed by near-field imaging (Fig. 1b), together with the extraordinarily low number of temporal modes, make this system extremely interesting for quantum optical metrology. Comparable sources of nonclassical light based on bulk crystals usually emit at least 400 modes [33].

In conclusion, noble-gas-filled hollow-core kagomé-PCF is ideal for creating ultrafast photon-number-correlated twin-beam SV. The central sideband frequency can be tuned over ~80 THz by simply changing the gas pressure. Technical limitations, like the dropping QE of the photodiode, prevented us from measuring noise reduction over the whole tuning range. However, there is no fundamental physical reason restricting the usable spectral tuning range of the source. The absence of SpRS means that photon-number correlated sidebands can be efficiently generated close to the pump wavelength, unlike in solid-core glass fibres. As a result, femtosecond signal and idler pulses can be generated along the whole length of the fibre, without significant group velocity walk-off. This opens up additional possibilities for ultrafast quantum optics using fibre sources. Furthermore, the very broad spectral bandwidth of ~50 THz makes the system highly interesting for spectroscopic applications. Also remarkable is the low ($< 5$) number of temporal modes of the spatial single mode system. The measured twin-beam squeezing of ~35% below the shot-noise level is mainly limited by the losses in the PCF itself and in the optical channels after the PCF. Photon-number correlation has been achieved for a squeezed vacuum state with a record brightness of ~2500 photons per mode. The system overcomes many of the material restrictions of solid-state systems and has the potential to push twin-beam generation into the UV. In the future, it should be possible to tailor the kagomé-PCF dispersion to the pump laser, permitting a further reduction in the number of temporal modes and allowing realization of a Raman-free single spatiotemporal mode source for in-fiber generation of factorable twin-beams.

## Acknowledgement

The research received partial financial support from the EU FP7 under grant agreement No. 308803 (project BRISQ2).

# Raman-free, noble-gas-filled PCF source for ultrafast, very bright twin-beam squeezed vacuum

## Supplementary Information

**Martin A. Finger[1*], Timur Sh. Iskhakov[1†], Nicolas Y. Joly[2,1], Maria V. Chekhova[1,2,3] and Philip St.J. Russell[1,2]**

[1]*Max Planck Institute for the Science of Light and* [2]*Department of Physics, University of Erlangen-Nuremberg, Guenther-Scharowsky Strasse 1, Bau 24, 91058 Erlangen, Germany*

[3]*Physics Department, Lomonosov Moscow State University, Moscow 119991, Russia*

*\*e-mail: martin.finger@mpl.mpg.de*

**S1. Derivation of the four-wave mixing Hamiltonian**

Starting from macroscopic electric fields we derive the expression for the four-wave mixing (FWM) Hamiltonian, in which the pump field is treated classically while the signal and idler fields are quantized. The Hamiltonian can be expressed as the volume integral of the material polarization multiplied by the electric field [1]. Formally, this means that the Hamiltonian for third-order nonlinear processes is proportional to the volume integral of the third-order electric susceptibility times the interacting fields,

$$H \propto \int d^3r \; \chi^{(3)} E^4 \,. \tag{1}$$

In the case of noble gas-filled hollow-core kagomé PCF it is valid to neglect the tensor nature of $\chi^{(3)}$ by assuming an isotropic dielectric medium. Moreover, since there is no Raman effect in the system, equation (1) assumes the nonlinear response of the medium to be instantaneous. In terms of complex electric fields, the field generated in the medium can be described as the pump field plus signal and idler fields and their complex conjugates, $E = E_p + E_s + E_i + c.c.$. When inserted into the Hamiltonian, this expression leads to a large number of terms describing all possible third-order nonlinear interactions. The term responsible for degenerate FWM can be identified to have the form $E_{FWM} = 4E_p^2 E_s^* E_i^* + c.c.$, which leads to the FWM Hamiltonian

$$H_{FWM} \propto \int dz\, \chi^{(3)} E_p^2 E_s^* E_i^* + c.c. \qquad (2)$$

with z being the coordinate along the fibre. Due to the high number of photons in the pump beam we can apply the parametric approximation and treat the pump as a classical field $E_p^2 = \int d\omega_1 E_{p1}(\omega_1) e^{-i\omega_1 t + i\beta_1 z} e^{-i\gamma_1 P_1 z} \int d\omega_2 E_{p2}(\omega_2) e^{-i\omega_2 t + i\beta_2 z} e^{-i\gamma_2 P_2 z}$ neglecting the pump depletion and quantum fluctuations. The terms $e^{-i\gamma_{1,2} P_{1,2} z}$ are included to account for the phase shift of the pump due to self-phase modulation (SPM) with $P_{1,2}$ being the pump peak powers and $\gamma_{1,2} = n_2 \omega_{1,2} / (A_{eff} c)$ the nonlinear coefficient [2, 3]. Here, $n_2$ is the nonlinear refractive index and $A_{eff}$ is the effective area. In the case of degenerate pump fields it holds that $\gamma_1 P_1 = \gamma_2 P_2 = \gamma P$. The signal and idler fields are treated as quantized. Passing to operators, the multimode fields can be written as $E_s^* \propto \int d\omega_s\, a_s^\dagger(\omega_s) e^{i\omega_s t - i\beta_s z}$ and $E_i^* \propto \int d\omega_i\, a_i^\dagger(\omega_i) e^{i\omega_i t - i\beta_i z}$, with $a_s^\dagger$, $a_i^\dagger$ being photon creation operators in the signal and idler beams [4]. This allows writing the Hamiltonian as

$$H_{FWM} \propto \chi^{(3)} \int dz \int d\omega_1 \int d\omega_2\, E_{p1}(\omega_1) E_{p2}(\omega_2) e^{-i(\omega_1+\omega_2)t} \\ \times \int d\omega_s \int d\omega_i\, a_s^\dagger(\omega_s) a_i^\dagger(\omega_i) e^{i(\omega_s+\omega_i)t} e^{i\Delta\beta z} + h.c. \tag{3}$$

with $\Delta\beta = \beta_1(\omega_1) + \beta_2(\omega_2) - \beta_s(\omega_s) - \beta_i(\omega_i) - \gamma_1(\omega_1) P_1 - \gamma_2(\omega_2) P_2$ defining the phase mismatch. Integrating (3) over the length $L$ of the fibre gives

$$H_{FWM} \propto \chi^{(3)} L \int d\omega_1 \int d\omega_2\, E_{p1}(\omega_1) E_{p2}(\omega_2) e^{-i(\omega_1+\omega_2)t} \\ \times \int d\omega_s \int d\omega_i\, a_s^\dagger(\omega_s) a_i^\dagger(\omega_i) e^{i(\omega_s+\omega_i)t} \operatorname{sinc}\left(\frac{\Delta\beta L}{2}\right) + h.c. \tag{4}$$

This expression can be written more compact by denoting the integral in the pump frequencies as a certain function playing the role of the time-dependent joint spectral amplitude,

$$\Pi(t,\omega_s,\omega_i) = \int d\omega_1 \int d\omega_2\, E_{p1}(\omega_1) E_{p2}(\omega_2) e^{-i(\omega_1+\omega_2-\omega_s-\omega_i)t} \operatorname{sinc}\left(\frac{\Delta\beta L}{2}\right), \tag{5}$$

then

$$H_{FWM} \propto \chi^{(3)} L \int d\omega_s \int d\omega_i\, \Pi(t,\omega_s,\omega_i)\, a_s^\dagger(\omega_s) a_i^\dagger(\omega_i) + h.c. \tag{6}$$

It can be seen that this Hamiltonian resembles the PDC version [1] with the major differences that in the case of FWM it depends quadratically on the pump amplitude and the pump SPM should be taken into account. For systems with low interaction strength, first-order perturbation theory can be applied to calculate the resulting state as

$$|\psi\rangle \approx |0,0\rangle + \frac{1}{i\hbar} \int H_{FWM}\, dt\, |0,0\rangle. \tag{7}$$

After time integration, the resulting joint spectral amplitude distribution can be defined as

$$F(\omega_s, \omega_i) = \int d\omega_1 E_{p1}(\omega_1) E_{p2}(\omega_s + \omega_i - \omega_1) \operatorname{sinc}\left(\frac{\Delta\beta L}{2}\right), \tag{8}$$

in agreement with the result of [5].

## S2. Power dependence of the sidebands

In the modulation instability regime, phase matching between the pump and the sidebands is strongly influenced by the effects of self- and cross-phase modulation of the pump beam. Therefore, perfect phase-matching is given by $\Delta\beta = 0 = 2\beta_p - \beta_s - \beta_i - 2\gamma P$ with $\beta_{p,s,i}$ being the wave vectors for the pump, signal, and idler radiation and $P$ being the pump peak powers. This leads to a power dependence of the spectral position of the sidebands and makes the sidebands move away from the pump with increasing pump power. Fig. 1 shows the MI spectrum for increasing average power at a constant pressure of 75 bar. It can be seen that with increasing pump power the sidebands broaden due to higher gain and shift away from the pump. The frequency difference between the sidebands increases by ~40 THz when the average pump power of 42 mW inside the fibre is doubled. In practice, the power-dependent shift of the sidebands has to be taken into account when a certain frequency band is targeted for generating twin beams.

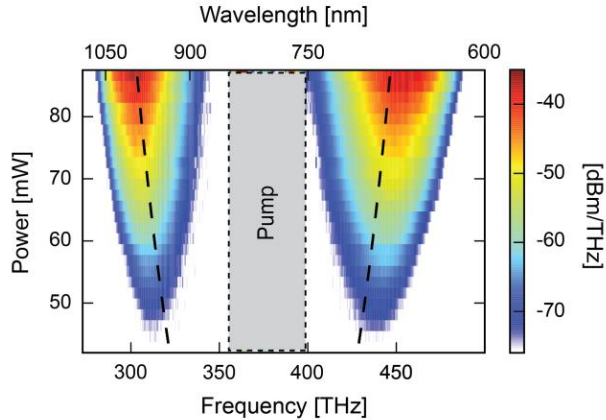

FIG. S1 (color). MI sidebands measured for a constant pressure of 75 bar and increasing average pump power. The pump was blocked with notch filters to avoid spectral artefacts from the OSA. The dashed line shows the calculated curve for perfect phase-matching based on an empirical kagomé dispersion model.

**S3. Calibration of the shot-noise level**

We have calibrated the shot-noise level of our detectors with the strongly attenuated output of the Ti:Sa amplifier [6]. For this, the laser beam was split on a polarizing beam splitter and adjusted so that both photodiodes measured nearly equal signals. Before the beam was sent to the photodiodes, all ambient light was blocked with two bandpass filters centered at 800 nm (bandwidth: 12 nm). Subsequently, we measured the power dependence of the variance of the difference signal of our detectors. This procedure was done for signal ratios of 0.9, 1 and 1.1 to make the calibration robust to small changes of the ratio. Fitting these dependences simultaneously with one function allows extracting a reliable boundary for the shot noise. Figure 2 shows the NRF measurement for the coherent state of the laser after the calibration.

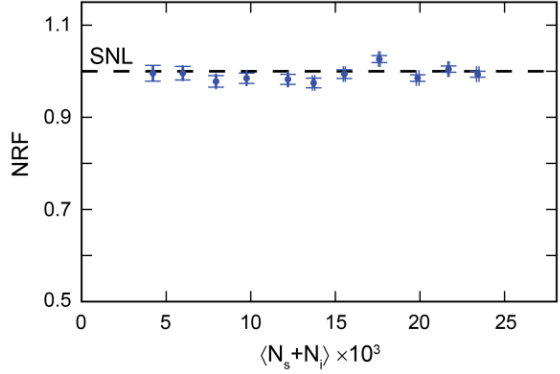

FIG. S2 (color). NRF measured for the Ti:Sa laser versus the sum of the mean photon numbers in both channels, varied by changing the laser power. Error bars represent the standard error of the measurement.

**S4. Estimation of losses in the optical channel**

In order to determine the best NRF achievable with our setup, we estimated the loss in the optical channel, starting from the kagomé PCF. Since the loss of the kagomé PCF was different for the two sidebands by ~3 dB/m, we were estimating the quantum efficiency of the high-loss sideband. In order to achieve noise reduction at high photon numbers, the losses in the two channels were balanced optically, by means of additional loss introduced in the low-loss sideband. The loss of the kagomé PCF was measured to be ~5 dB/m for the high-loss (long-wavelength) sideband and therefore led to a transmission of $T_{PCF} = 0.71$ for the used 30 cm fibre. Next, the sidebands passed through the fused silica glass window of the gas cells with a transmission of $T_{window} = 0.94$. The achromat, half-wave plates (HWP) and the focussing lenses in front of detectors were all antireflection coated with an estimated transmission of $T_{AR} = 0.99$. Also the Brewster angle prism transmission $T_{Brewster} = 0.99$ was taken into account. The notch filters (NF) had a transmission of $T_{notch} = 0.98$ and the polarising beam splitters (PBS) had $T_{PBS}$

= 0.95. The transmission values for the long-pass (LP) and short-pass (SP) filters were assumed to be $T_{LP/SP} = 0.97$. The reflection of the dielectric mirrors was $R_{dm} = 0.99$ and of the silver mirrors $R_{sm} = 0.96$. The quantum efficiency (QE) of the detectors was 0.95. It was assumed to be constant over the wavelength range of interest, since it varied only by ±2%. This amounted to a total quantum efficiency (transmission) of the whole optical channel of 0.45 and set the limit for the best achievable $NRF_{best} = 0.55$. This value was in reasonable agreement with the measured NRF of ~0.65. The small discrepancy can be explained with the frequency-dependent loss of all components and especially of the kagomé PCF. The given transmission values corresponded to an estimated average over the wavelengths range of interest.

## S5. Derivation of the relation between $g^{(2)}$ and the effective number of modes

We are presenting here the derivation of the relation between the second-order intensity autocorrelation function $g^{(2)}$ and the effective number of modes given by the Schmidt number [7]. The Schmidt number is calculated as $K = 1 / \sum_i \lambda_i^2$, where $\lambda_i$ are the Schmidt eigenvalues.

For a twin-beam squeezed vacuum, the eigenvalues are found from the Schmidt decomposition of the two-photon amplitude [8]. A single (signal or idler) twin beam taken separately will be a mixture of independent thermal modes with different intensities, determining their weights in the mixture. The weights coincide with the Schmidt eigenvalues [8]. The total intensity is

$$I = \sum_i \lambda_i I_i. \qquad (9)$$

Plugging this into the definition of $g^{(2)}$ leads to

$$g^{(2)} = \frac{\left\langle \left(\sum_i \lambda_i I_i\right)^2 \right\rangle}{\left(\sum_i \lambda_i \langle I_i \rangle\right)^2} = \frac{\sum_i \lambda_i^2 \langle I_i^2 \rangle + \sum_{i \neq j} \lambda_i \lambda_j \langle I_i I_j \rangle}{I^2}, \quad (10)$$

with $\langle I_i \rangle = I$. Since the modes are independent, it holds that $\langle I_i I_j \rangle = I^2(1 + \delta_{ij})$. This leads to the expression

$$g^{(2)} = \frac{2I^2 \sum_i \lambda_i^2 + I^2 \sum_{i \neq j} \lambda_i \lambda_j}{I^2} = 2\sum_i \lambda_i^2 + \sum_{i \neq j} \lambda_i \lambda_j. \quad (11)$$

Remembering that $\sum_i \lambda_i^2 + \sum_{i \neq j} \lambda_i \lambda_j = \left(\sum_i \lambda_i\right)^2 = 1$, allows one to find the final form of the equation

$$g^{(2)} = 1 + \sum_i \lambda_i^2 = 1 + \frac{1}{K}. \quad (12)$$